\documentclass[aps,prl,referee]{revtex4-1}

\usepackage{xspace}
\newcommand{\ket}[1]{\ensuremath{|#1\rangle}\xspace}

\usepackage{graphicx}
\usepackage{dcolumn}
\usepackage{bm}
\usepackage{color}

\begin{document}

\title{A cold atom pyramidal gravimeter with a single laser beam }

\author{Q. Bodart,$^1$ S. Merlet,$^1$ N. Malossi,$^1$ F. Pereira Dos Santos,$^1$ P. Bouyer,$^2$ and A. Landragin$^1$}

\email{arnaud.landragin@obspm.fr}

\affiliation{$^1$LNE-SYRTE, Observatoire de Paris, CNRS, UPMC, 61
avenue de l'Observatoire, 75014 Paris, France}

\affiliation{$^2$Laboratoire Charles Fabry de l'Institut
d'Optique, CNRS et Univ Paris Sud, Campus Polytechnique, RD 128,
91127 Palaiseau Cedex, France}

\date{15 December 2009}

\begin{abstract}
We demonstrate a scheme for realizing a compact cold atom gravimeter. The use of a hollow pyramidal configuration allows
to achieve all functions: trapping, interferometer  and detection with a unique laser beam leading to a drastic reduction in complexity and volume. In particular, we demonstrate a relative sensitivity to acceleration of gravity (g) of $1.7 \times 10^{-7}$ at one second, with a moderate laser power of 50~mW. This simple geometry combined to such a high sensitivity opens  wide perspectives for practical applications (P. Bouyer and A. Landragin, patent n $^{\circ}$ FR2009/000252, 2009).
\end{abstract}

\maketitle

Gravimeters, based on atoms interferometry, measure the Earth's gravity as a phase shift between two paths of matter waves~\cite{Peters}. Applications of such sensitive gravimeters cover numerous fields, from fundamental physics~\cite{balancew,fixler,lamporesi,Mueller08} to navigation and geophysics. However transportable interferometers are required for foreseen applications in the field of navigation and gravity field mapping~\cite{Bell}. In this paper, we show that an atom interferometer based on Raman transitions~\cite{Kasevich1} can be realized exploiting an hollow pyramid. This geometry enables a drastic reduction in complexity of atomic gravimeters, replacing all laser beams (typically nine independent beams~\cite{Peters, Le Gouet}) by only one. Its permits building much more compact instruments with moderate laser power and with performances comparable to state of the art sensors. Pyramidal magneto-optical traps~(MOT)~\cite{Lee} have been introduced to reduce the size and simplify cold atoms experiments since they require a single laser beam. They have been used for different kinds of alkali atoms in various experiments, with a hole at the vertex for generating continuous beams~\cite{Kohel,Camposeo,Arlt}, for direct loading of 3D MOTs~\cite{Williamson}, for quantum gases experiments~\cite{Vangeleyn} and eventually in microscopic pyramids on atom chips~\cite{Pollock}. In our experiment, the many reflections of the single incident beam onto the four mirrors of the hollow pyramid allow to obtain the needed polarizations not only for trapping the atoms inside the pyramid but also for efficiently driving the Raman transitions and for detecting the atoms~\cite{brevet}.

\begin{figure}[h]
       \includegraphics[width=9.5cm]{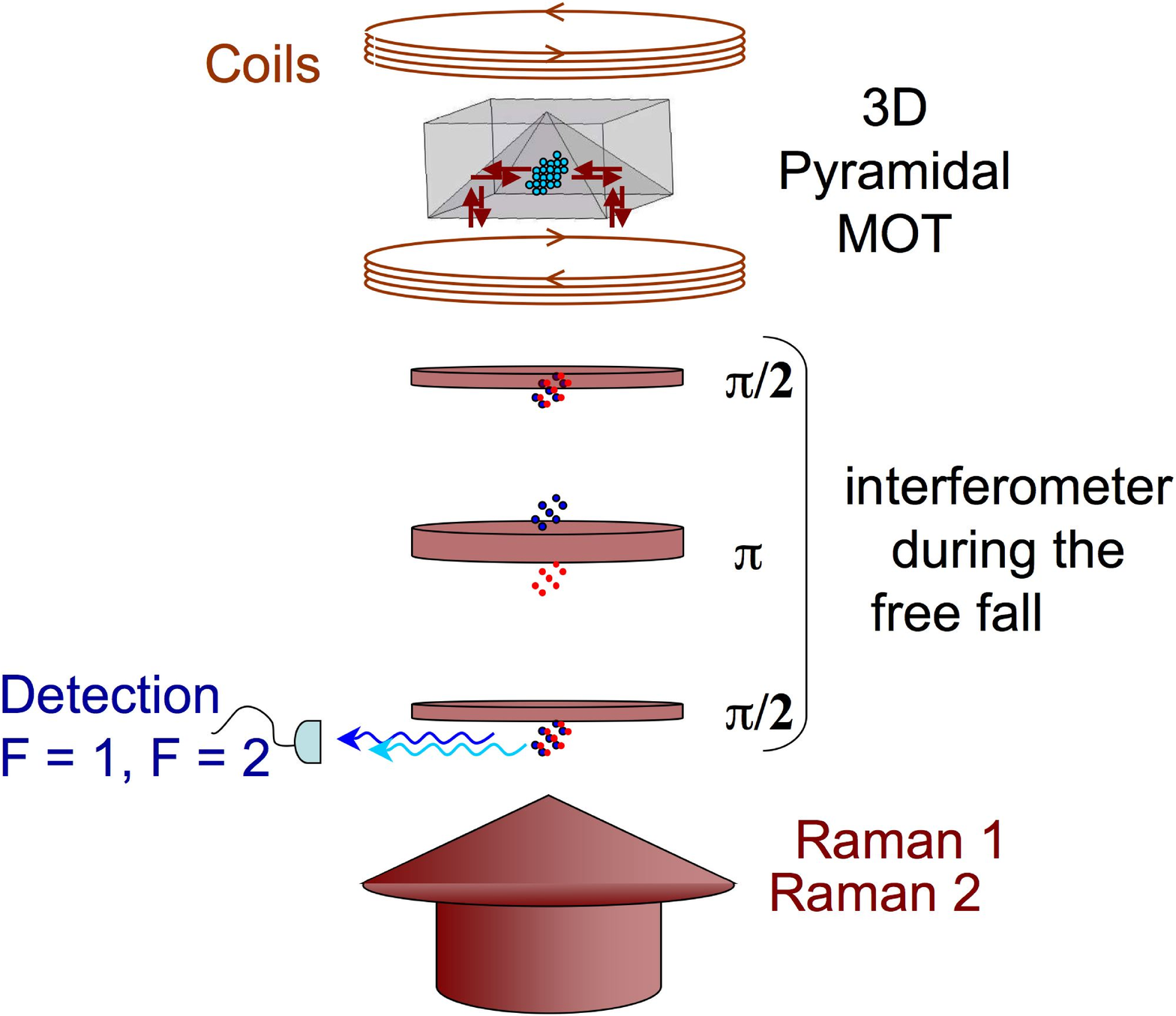}
    \caption{Experimental setup of the experiment; the two frequency collimated laser beam arrives from the bottom by one single fiber and cools down $1.2~10{^6}$ atoms to
    $2.5~\mu$K. At each measurement cycle, these two laser frequencies are detuned
    to realize the Raman pulses of the interferometer. The total high from the top of the pyramid to the detection is of 15 cm.}
    \label{experiset}
\end{figure}

We first briefly describe the experimental setup, pictured in Fig.~\ref{experiset}, and the principle of operation of the
gravimeter. We load a magneto optical trap directly from a vapor
of $3 \times 10^{-9}$ hPa of Rubidium 87 (Rb). At each cycle, the
pyramidal MOT traps typically $4 \times 10{^6}$ atoms in $360$~ms.
After a molasses stage of 20~ms, we switch off the light
adiabatically and let the atoms fall. We have measured by Raman
velocimetry~\cite{Kasevich2} a temperature of $2.5~\mu$K, as low
as the one obtained with the usual configuration based on six
independent laser beams. In order to reduce the sensitivity to
magnetic field, we select atoms in the~$\ket{F=1,m_{F}=0}$ state,
using a sequence of a micro-wave and pusher beam pulse at the
beginning of the free fall. As soon as the atoms have left the
pyramid, we perform a velocity selection in the vertical
direction, leaving about $3\times10{^5}$ cold atoms within state
$\ket{F=1,m_{F}=0}$. Then we realize an interferometer with a
$\pi/2-\pi-\pi/2$ sequence~\cite{Kasevich1}, with an interrogation
time of up to 80~ms. Thank to the internal state labeling technique ~\cite{Borde}, the interferometer phase shift is extracted from the measurement of the population in both output ports of the interferometer by laser-induced fluorescence on their associated internal states $\ket{5S_{1/2},F=1}$ and $\ket{5S_{1/2},F=2}$. The total sequence duration lasts 560~ms. In addition, the entire experiment is placed onto a passive isolation platform in order to reduce the influence of spurious vibrations.

The use of a single laser beam leads to many simplifications in the laser system as it avoids all optical elements for the splitting, transport and independent power control of many beams. This single beam is composed of two frequencies, whose difference corresponds to the microwave transition of the Rb ground levels (i.e. 6.8 GHz). The laser system consists in a simplified version of the one already described in detail in~\cite{Cheinet}. We briefly recall here the main features. Two extended-cavity diode lasers are tuned respectively close to the $\ket{5S_{1/2},F=1}\rightarrow \ket{5P_{3/2}}$ and $\ket{5S_{1/2},F=2}\rightarrow \ket{5P_{3/2}}$ transitions at 780~nm. The laser frequencies are independently controlled thanks to a versatile electronic system, based on beatnote measurements and frequency to voltage conversion. This system allows fast changes of the laser frequencies, from very close to resonance for the cooling phase to detuned by 151 MHz with respect to $\ket{5P_{3/2},F^{'}=1}$ for the interferometer. These two laser beams are superimposed onto a polarizing beamsplitter cube, and finally injected into the same polarization-maintaining fiber with a 1:4 ratio. At the output of the fiber, the beam has a total power of 50~mW and is circularly polarized. It is collimated with a waist of 14~mm ($1/e^2$ radius) before entering the vacuum chamber from the bottom.

The inverted pyramid was manufactured~\cite{fichou} out of two glass corner cubes and two glass isosceles rectangular prisms glued together, on a pedestal. The corner cubes are glued on the four ridges in order to obtain right angles between the opposite faces inside the pyramid. We have controlled the angle to be $90^{\circ}$ within one arc minute. The pyramid base is a square of $20 \times 20~$mm$^2$ area. The pyramid is attached to the upper flange pointing upward. Inner faces are dielectrically coated for maximum reflection at $45^{\circ}$ and for equal phase-shift between the two orthogonal polarizations. Therefore, reflections of the single beam onto the four inner faces of the pyramid creates the required
3D $\sigma^+$/$\sigma^-$ polarization configuration for the trap inside the pyramid~\cite{Lee}. In addition, outside the pyramid, the laser field configuration allows driving either
$\sigma^+$/$\sigma^+$ or $\sigma^-$/$\sigma^-$ vertical velocity selecting Raman transitions~\cite{Kasevich1} needed to realize the interferometer and the detection.

\begin{figure}[h]
       \includegraphics[width=8cm]{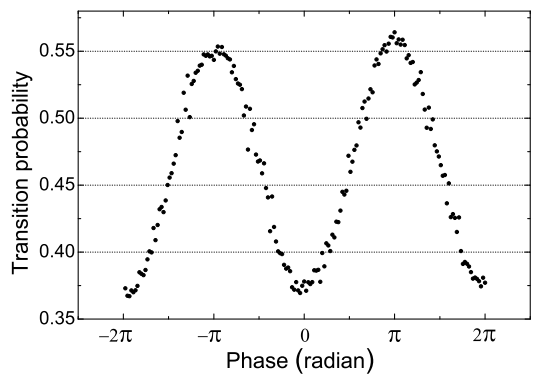}
    \caption{Interferometer fringe pattern for a total interferometer time of 80 ms. Each point corresponds to a single measurement.}
    \label{transprob}
\end{figure}

We now characterize the performances of the interferometer using the pyramid. The measurement of g is extracted from the frequency chirp of the Raman laser difference that compensates exactly the time-dependent Doppler shift, as described in~\cite{Cheinet}. Figure~\ref{transprob} displays the fringe pattern obtained for the maximal total interferometer duration of 80 ms and a Raman pulses sequence $\tau-2\tau-\tau$, with $\tau=9~\mu$s. The fringe pattern is obtained by scanning the transition probability between hyperfine quantum states versus the interferometer phase-shift induced by a phase jump on the Raman lasers phase difference between the second and the third pulse. The vibration noise is reduced by using corrections obtained from the signal of a highly sensitive seismometer~\cite{Le Gouet}. We obtain a contrast of 19~\% and a signal to noise ratio of 23 shot to shot at half fringe.

\begin{figure}[h]
       \includegraphics[width=7.5cm]{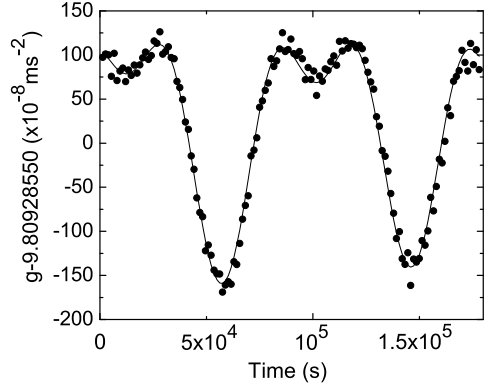}
    \caption{Variation of Earth's gravity as a function of the time. The dots present experimental data averaged over 22 minutes. The line corresponds to predicted Earth's tides.}
    \label{marees}
\end{figure}

Continuous gravity measurements have been recorded during 50 hours and agree with the results of predicted Earth's tides~\cite{PETGTAB}, as displayed on Figure~\ref{marees}. In order to track fluctuations of gravity, we servo-lock the value of the frequency chirp to compensate for the Doppler effect at any time. The error signal of the servo-loop is calculated from the difference of alternative measurements on the two sides of the central fringe.

\begin{figure}[h]
       \includegraphics[width=8cm]{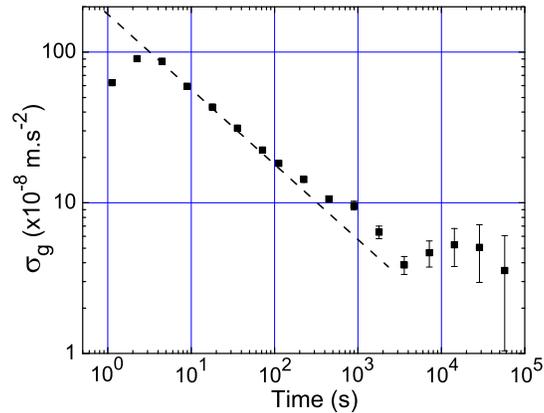}
    \caption{Allan Standard deviation of relative gravity measurements corrected from the Earth's tides.}
    \label{Allan}
\end{figure}

Figure~\ref{Allan} shows the Allan standard deviation of the relative g measurements when subtracting the tides signal calculated from the model. The sensitivity improves as $1.7 \times 10^{-6}$~m.s$^{-2}.\tau^{-1/2}$ up to 1000 s, where $\tau$ is the measurement time. This correspond to a relative short term sensitivity to g at one second of $1.7 \times 10^{-7}$). This performance lies about one order of magnitude above best atomic gravimeters~\cite{Mueller08,Le Gouet}. The technical limits were attributed to residual vibrations on the platform on one hand and spurious phase noises on the micro-wave reference frequency used for the Raman transitions on the other hand. Long term stability reaches  a flicker floor below $5 \times 10^{-9}$ g. Its limit is attributed to fluctuations of the systematic error corresponding to Raman laser wave-front distortions.

The wavefront distortion from the reflections leads to a systematic effect on the interferometer phase. We evaluate this phase shift as a function of the transverse expansion of the cloud by changing the temperature during the molasses phase. We found a linear dependance of about $5 \times 10^{-7}$ m.s$^{-2}$ per $\mu$K. Finally, we have also estimated the modification of the effective wave-vector of the Raman transition by the Rb background pressure compared to perfect vacuum, which leads to a systematic error of the level of $6 \times 10^{-8}$~m.s$^{-2}$.

These performances, which have been obtained despite a moderate contrast, can be improved in many ways.
Higher atom number, which enables a thinner velocity class selection by the Raman transition, can be achieved with higher laser power. This will also allow for a larger laser beam size and Raman detuning, thus reducing transverse laser intensity variations and the decoherence by spontaneous emission (3.6\% in this experiment). Improvement of the pyramid reflector will reduce the intensity and phase inhomogeneities across the beam. Because of the radial expansion of the cloud, the transverse motion of the atoms or a change in their initial position, these homogeneities result in a reduction of the contrast, from 31\% for 2T = 2 ms to 19\% for 2T = 80 ms. In our experiment, the better contrast was obtained when atoms were located far from the dark lines corresponding to the edges of the pyramid faces.

In conclusion, we demonstrated that a pyramidal reflector allows to realize a compact atomic gravimeter with only one laser beam of moderate power. The short term sensitivity was limited by technical noises, which can be reduced to  $6\times10^{-8}~$g.$\tau^{-1/2}$, as demonstrated in another experiment with similar parameters (interrogation time, vibration level of the ground, repetition rate, number of atoms)~\cite{gyro2009}. The fundamental limit, which is due to the atomic shot noise, has been estimated to be $3
\times 10^{-8}$~g.$\tau^{-1/2}$, and can be pushed down with larger laser power. Compared to commercial ballistic corner cube gravimeters, these performances are already comparable with compact models~\cite{A10}, and can reach performances at the level of the state of the art~\cite{FG5}.

The reduction in the complexity leads to a drastic reduction of the volume of the physical package from few hundred to few liters. As it also simplifies the optical bench and requires only a moderate optical power, similar reduction of the size of the optical part is also expected compared to these of standard cold atom experiments (for example by using  an all fibered system~\cite{Nyman}). Moreover, this gravimeter is scalable: increasing or decreasing the size of the pyramid and the height of free fall allows to change in the same time the expected performances by changing the number of collected atoms and the interrogation time. Such a system, combined with a compensation of the residual vibrations by correlation with the signal of a mechanical seismometer~\cite{Merlet}, opens the way for practical field applications in gravimetry and more generally in inertial force measurements with atom interferometry.

\begin{acknowledgments}
We would like to thank the Institut Francilien pour la Recherche
sur les Atomes Froids (IFRAF) and the European Union (EuroQUASAR/IQS project)
for financial supports. Q. B. thanks the CNES for supporting his work.
\end{acknowledgments}


\begin{thebibliography}{10}
\bibitem{fichou} The pyramidal reflector have been realized by Optique de Pr\'{e}cision J.Fichou, 30 rue de la Garenne, 94260 Fresnes, France.
\bibitem{PETGTAB} The model was carried out by applying the PETGTAB software, version 3.01, produced by G. Wenzel, Institut f\"{u}r Erdmessung, University of Hannover, Germany.
\end{thebibliography}
\end{document}